\documentclass{acmsiggraph}

\usepackage{amsmath,amssymb}   
\usepackage{amsfonts}
\usepackage{units}
\usepackage[dvipsnames]{xcolor}
\usepackage{microtype} 
\usepackage{wrapfig}
\usepackage{dsfont}

\newenvironment{tight_enumerate}{
        \begin{enumerate}
                \setlength{\itemsep}{1pt}
                \setlength{\parskip}{0pt}
                \setlength{\parsep}{0pt}
        }{\end{enumerate}}
\newenvironment{tight_itemize}{
        \begin{itemize}
                \setlength{\itemsep}{1pt}
                \setlength{\parskip}{0pt}
                \setlength{\parsep}{0pt}
        }{\end{itemize}}

\usepackage{booktabs}

\usepackage{titlesec}
\titlespacing{\paragraph}{%
        0pt}{
        0.05\baselineskip}{
        1em}%
\newcommand{\secref}[1]{Sec.~\ref{#1}}
\newcommand{\figref}[1]{Fig.~\ref{#1}}

\DeclareMathOperator*{\argmin}{argmin}

\newcommand{\X}{\mathbf{X}}

\newcommand{\x}{\mathbf{x}} 
\newcommand{\xl}{\x^\text{L}}
\newcommand{\xr}{\x^\text{R}}
\newcommand{\yt}{y^\text{Top}}
\newcommand{\yb}{y^\text{Bot}}
\renewcommand{\S}{S}
\newcommand{\C}{\mathcal{C}}
\newcommand{\Ci}{\C_i}
\newcommand{\Cyll}{\mathrm{Cyl}}
\newcommand{\Cyl}[1]{\Cyll(#1)}
\newcommand{\Strr}{\mathrm{Str}}
\newcommand{\Str}[1]{\Strr(#1)}
\newcommand{\Intt}{\mathrm{Int}}
\newcommand{\Int}[2]{\Intt(#1,#2)}

\usepackage{siunitx}
\usepackage[mathletters]{ucs}
\usepackage[utf8x]{inputenc}
\usepackage{upgreek}
\usepackage{dblfloatfix}

\usepackage{hyperref}
\hypersetup{
    pdfinfo={
        Title={Shape Representation by Zippable Ribbons},
        Author={Christian Sch\"uller, Roi Poranne, Olga Sorkine-Hornung},
        Subject={},
        Keywords={}
    }
}

\usepackage{t1enc,dfadobe}

\title{
Shape Representation by Zippable Ribbons
}

\author{
Christian Sch\"uller\\
ETH Z\"urich
\and
Roi Poranne\\
ETH Z\"urich
\and
Olga Sorkine-Hornung\\
ETH Z\"urich
}
\pdfauthor{Christian Sch\"uller, Roi Poranne, Olga Sorkine-Hornung}

\begin{document}

\teaser{
\includegraphics[width=1\textwidth]{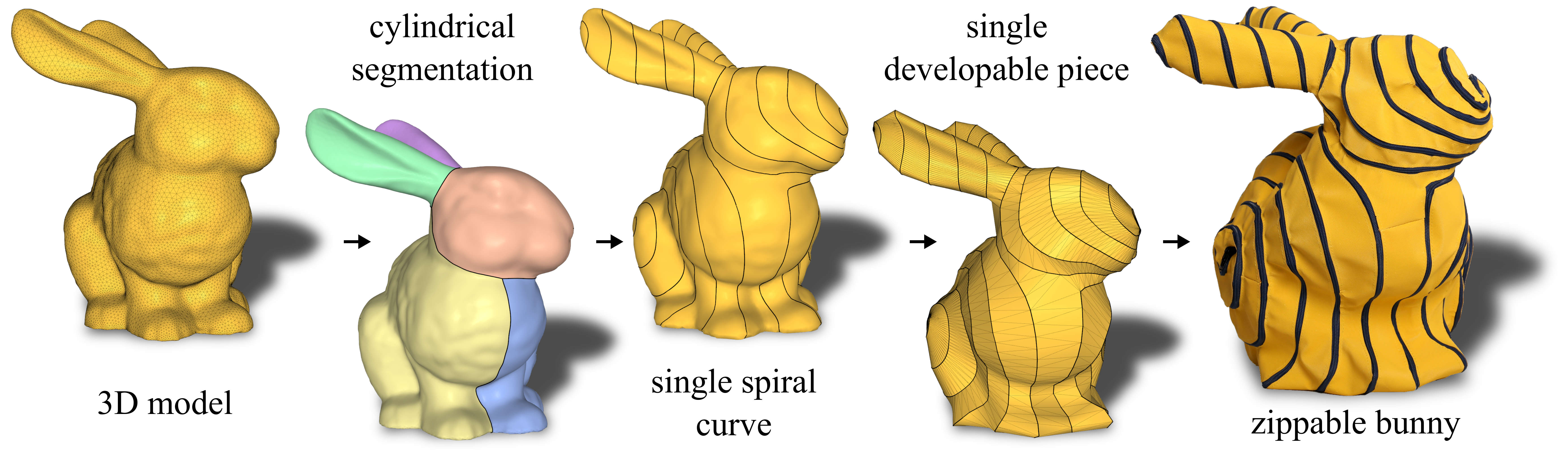}
\caption{The pipeline of our approach. Starting from a 3D model the user segments the shape into topological cylinders. Our algorithm then produces a curve on the shape that spirals along the cylinders. It proceeds to cut the shape along the curve and creates a single, developable ribbon-like surface. The ribbon is flattened into 2D and, based on the flattening, plans for laser cutting the ribbon out of fabric are generated. Finally, we stitch a single zipper to the boundary of the ribbon. Zipping it up reproduces the original shape.}
\label{fig:teaser}
}

\maketitle

\begin{abstract}
Shape fabrication from developable parts is the basis for arts such as papercraft and needlework, as well as modern architecture and CAD in general, and it has inspired much research. We observe that the assembly of complex 3D shapes created by existing methods often requires first fabricating many small flat parts and then carefully following instructions to assemble them together. Despite its significance, this error prone and tedious process is generally neglected in the discussion. We propose an approach for shape representation through a \emph{single} developable part that attaches to itself and requires no assembly instructions. Our inspiration comes from the so-called \emph{zipit} bags \cite{zipit}, which are made of a single, long ribbon with a zipper around its boundary. In order to ``assemble'' the bag, one simply needs to zip up the ribbon. Our method operates in the same fashion, but it can be used to approximate \emph{any} shape. Given a 3D model, our algorithm produces plans for a single 2D shape that can be laser cut in few parts from flat fabric or paper. We can then attach a zipper along the boundary for quick assembly and disassembly, or apply more traditional approaches, such as gluing and stitching. We show physical and virtual results that demonstrate the capabilities of our method and the ease with which shapes can be assembled.

\end{abstract}


\section{Introduction}

\begin{figure}[b]
        \centering
        \includegraphics[width=1.0\columnwidth]{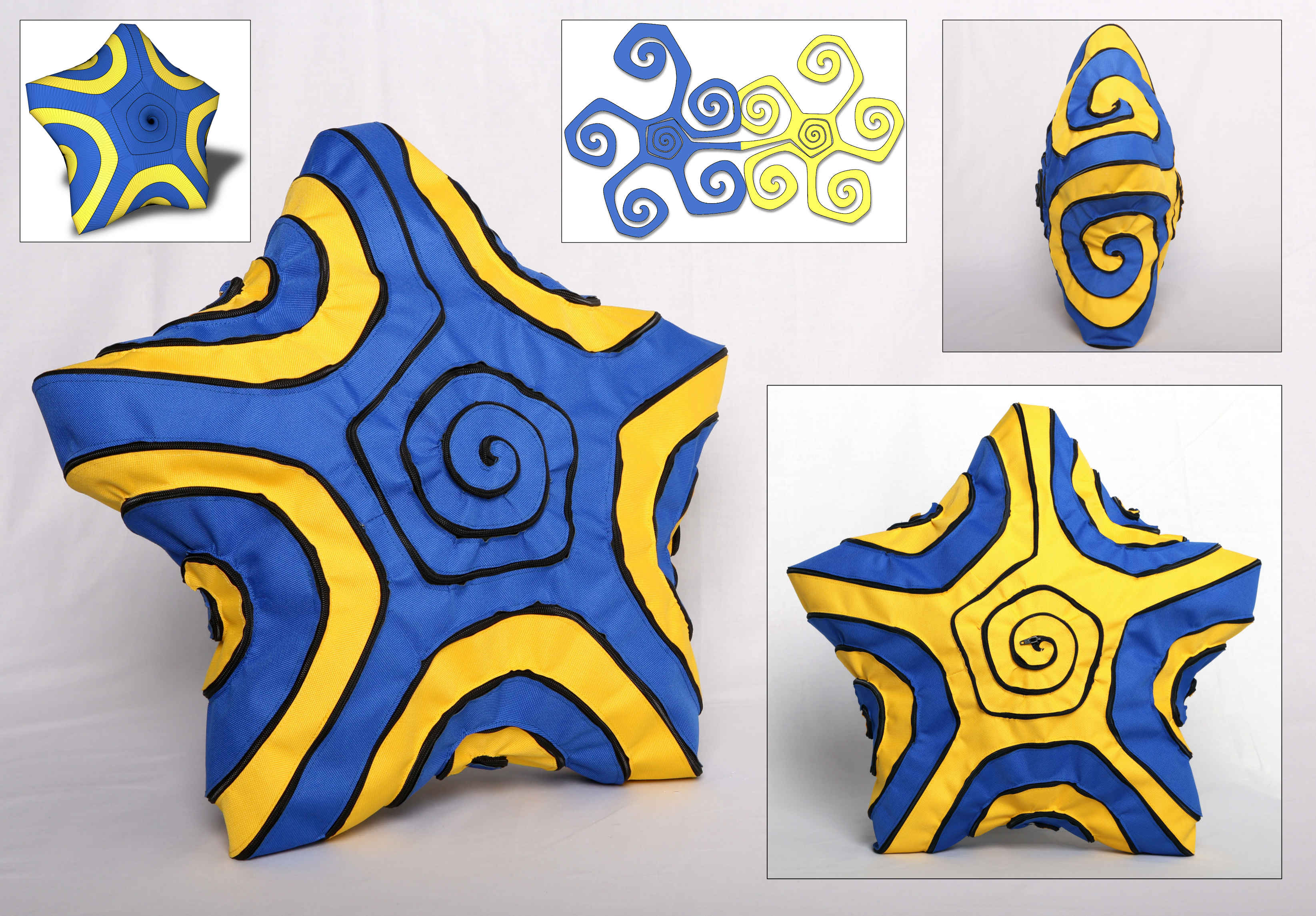}
        \caption{Our design for a zippable star pillow, made of two differently colored fabrics, flatly attached together.}
        \label{fig:Star}
\end{figure}

Representing shapes using developable surfaces is a problem with numerous applications, ranging from recreational activities like papercraft fabrication, to large scale industrial design and modern architecture. Our interest in this problem is inspired by a product commercially known as the \emph{zipit} bag \cite{zipit}. This bag is made from a single, long piece of fabric ribbon with a zipper attached around its boundary. When zipped up, the ribbon wraps around to form a simple bag. It takes only a few seconds to assemble and disassemble it, and no instructions are necessary. The simplicity of this concept and the oddly satisfying sensation of seeing it taking form when being zipped up immediately propelled us to ask whether this idea can be generalized to \emph{arbitrary} shapes. To this end, we devise a computational method to create a suitable developable shape that approximates a given target 3D model when zipped up; see e.g.\ \figref{fig:teaser}. Our approach generalizes the simple straight ribbons that make up the \emph{zipit} bags by allowing the ribbons to turn, have varying width, and {branch} (see \figref{fig:Star}).
Merely finding a 2D shape that approximately reproduces a 3D shape when zipped up is not very challenging. One approach would be to take a 3D mesh and cut it into a single triangle strip \cite{Rossignac99,EppsteinG04}. However, in addition to being mesh dependent, the resulting strip would have many sharp turns that make zipping up difficult, if not impossible, and the strip's width would be uneven, which can be visually unappealing. Therefore, we postulate the following desirable ``regularity'' properties for a zippable 2D ribbon shape:
\begin{tight_itemize}
\item The ribbon should curve as little as possible.
\item It should have as constant as possible width.
\end{tight_itemize}
The challenge we aim to solve is to compute a single, flat piece that satisfies these two properties and approximates the input shape well when zipped up.
We observe that the two properties are nearly trivial to achieve when the target 3D model is a cylinder: it is easy to trace a spiraling curve on a cylindrical surface such that if cut along that curve, the resulting shape is a straight ribbon with fixed width. 

For general target surfaces, our approach is based on first decomposing the shape into topological cylinders, and then mapping them onto cylindrical domains with low isometric distortion and in a seamless manner. We then draw ``perfect'' spirals on the cylinders and map these spirals back onto the input shape. 
\setlength{\intextsep}{8pt}%
\setlength{\columnsep}{8pt}%
\begin{wrapfigure}{r}{0.27\columnwidth}
        \centering
        \includegraphics[width=\linewidth]{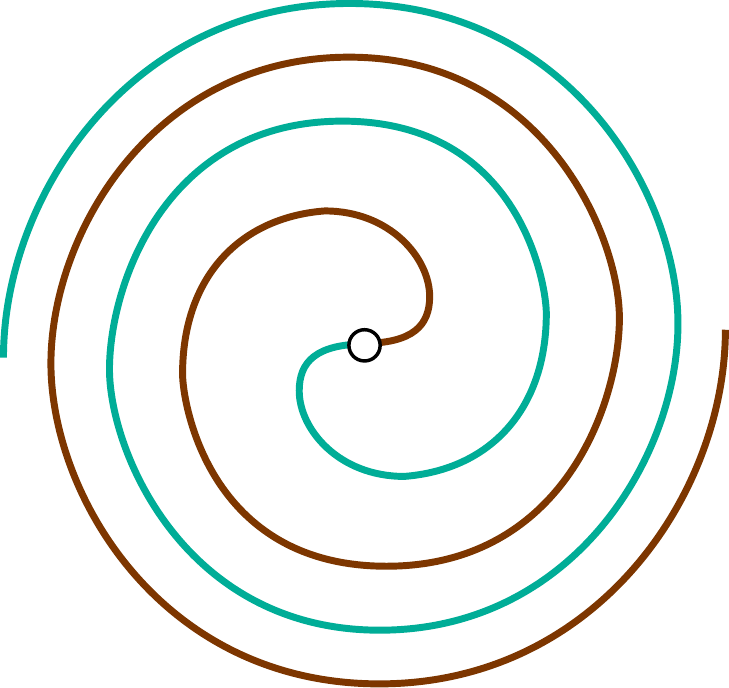}
\end{wrapfigure}
Since the mappings minimize isometric distortion, the mapped curves tend to exhibit low curvature and low variation of the distance between windings. 
Inspired by \cite{ZhaoGHGCTBZCC16} we connect the spirals on the different cylinders into a single, long \emph{Fermat spiral} (see inset). The shape is then cut along the computed curve to create a single, possibly bifurcated, but not yet developable ribbon. A simple remeshing process then transforms this ribbon into a developable one, which can be trivially unfolded onto the 2D plane to create the cutting pattern. It is possible that some strands of the flat ribbon overlap in the plane, and the pattern might take up too much space to fully fit onto the laser cutter bed; in both cases we simply divide the ribbon into a few separate pieces that are later sewn back together, before attaching the zipper (see Fig. \ref{fig:Pipeline}). 

In addition to the assembly process being easy and entertaining, our fabrication process has distinct advantages over papercraft and many other similar methods in this domain. Our assembly is \emph{linear}, i.e., at every instant of the assembly process, the next action is unique and unambiguous, and it requires \emph{no instructions}. In contrast, papercraft and related approaches typically require certain assembly order, especially for the final stages, where all the parts have to come together for the shape to close up; the makers usually must refer to a manual, find the next piece and understand how to attach it to their work. In case our method generates self-overlapping strands of ribbon that must be severed in order to laser-cut them from fabric, sewing the pieces together is simple since both ends are flat, perfectly matching in length and only require flat stitching along a straight line. Also sewing the zipper to the ribbon only requires working on flat sheets of fabric and is therefore straightforward and rather fast, as confirmed by the seamsters and seamstresses we employed. In contrast, attaching multiple parts in papercraft or sewing plush toys from multiple charts usually cannot be done in a flat position, and may necessitate more time and skill for large models.  

\begin{figure}[h]
        \centering
        \includegraphics[width=1.0\columnwidth]{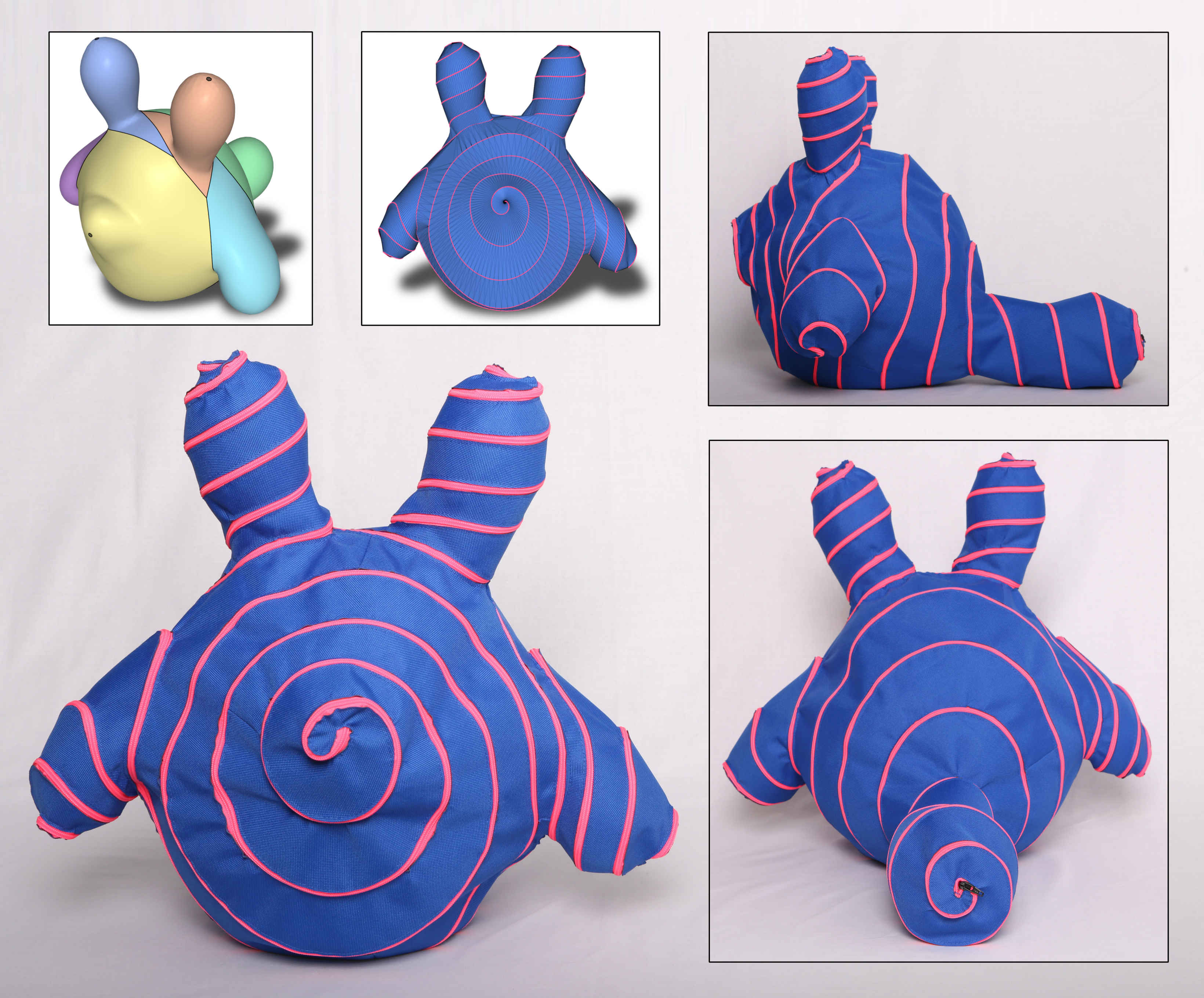}
        \caption{A zippable model of a  character. The zipper starts at the tail and spirals around all extruding parts until it ends at the tip of the nose.}
        \label{fig:Totoro}
\end{figure}

We demonstrate our technique on a number of various shapes, see e.g.\ the model in \figref{fig:Totoro}. We show virtual results and physically fabricated objects, which are  assembled and disassembled simply by slider a zipper. 

\section{Related work}\label{Sec:related}
Our work relates to the general field of computational fabrication and digital geometry processing. We briefly review the most relevant previous works below.

\paragraph{Papercraft and needlework.}
Some objects, typically of limited size and/or with certain constraints on the shape, can be directly manufactured in one piece using e.g.\ 3D printing or CNC milling. However, a large number of approaches propose to create shapes composed of several individually fabricated parts. In this space, the most related approaches to ours are papercraft based on cutting and gluing \cite{MitaniS04,ShatzTL06,MassarwiGE07,straub2011creating,TakahashiWSLY11}, and manufacturing by sewing~\cite{julius2005d,MoriI07,IgarashiI08,IgarashiIS09,Wang10,mahdavi2014cover}. Both types of methods require the approximation of a given 3D shape by pieces of developable surfaces (in the case of paper) or highly stretch-resistant material (in the case of fabric).
This is typically achieved by segmenting or cutting the shape into parts with low Gaussian curvature and parameterizing each part onto the plane. Alternatively, the shape can be a priori modeled or approximated as a piecewise developable surface, which is a current topic of active research~\cite{KolmanicG00,Rose:DevSurfaces:2007,KilianFCMSP08,LiuLH09,ZengLCY12,ChandraKKSAKW15,Tang:2016}. The assembly by gluing or sewing the pieces together requires precision and carefully following the instructions. In contrast, in our case, it is mostly a straightforward and even mindless task.

\paragraph{Soft materials.} 
Designing for fabrication using soft material like fabric or rubber is challenging, since such materials easily deform under stress and gravity. The main goal is to predict the deformation and solve an inverse problem, so that the fabricated model assumes the desired shape when subjected to the stress. Examples include the generation of plush toys \cite{MoriI07,IgarashiI08}, inflatable structures \cite{SkourasTBG12,Grinspun14} and rubber objects~\cite{Bickel:2010:DFM,Skouras:2013,chen2014anm}, among others. As mentioned, although we use fabric in our results, the behavior is more reminiscent of papercraft, because we use a nearly inextensible cloth and the zipper itself is rather stiff and completely inextensible.
Regardless of the type of fabric used, our zipper based approach has one decisive advantage: The whole fabrication process can be done entirely in the flat. In contrast, previous methods result in spaciously curved pieces once they get connected to each other, which makes sewing or gluing much harder as confirmed by professional tailors. The final assembly by zipping up needs no instructions, is fast and fun to do.

\paragraph{Parameterization.}
Our approach relies on mesh parameterization, which is an extensively studied topic, see e.g.\ the survey in \cite{HormannLS07}. The more recent relevant parameterization literature is presented in \cite{Smith:2015,Kovalsky:2016,Rabinovich:SLIM:2017}.
Our method introduces a \emph{new type} of global parameterization, which extends cylindrical parameterization~\cite{Tarini12}. Global parameterization is primarily used for quad meshing, where parts of seams in the parameter domain are related to each other by a rotation of integer multiples of $\pi/2$. A recent review can be found in \cite{QuadMeshingSTAR:2013}. 
In our specific case, we require a different form of seamless mapping, based on cylindrical domains~\cite{RayLLSA06,KnoppelCPS15}. Our main inspiration is \cite{kalberer2011stripe}, where the authors propose an approach for drawing stripes on \emph{tubular} shapes that can be used for generating textures. Their approach is based on a seamless parameterization aligned with a 2-RoSy field obtained from the principal curvature directions. This causes problems near umbilical points, where the principal directions are unstable. In contrast, we employ parameterization based on distortion minimization, which avoids this problem, and generally leads to less distorted mappings \cite{MylesZ12}.

\section{Method}
\label{sec:Method}

\begin{figure}[b]
   \includegraphics[width=\columnwidth]{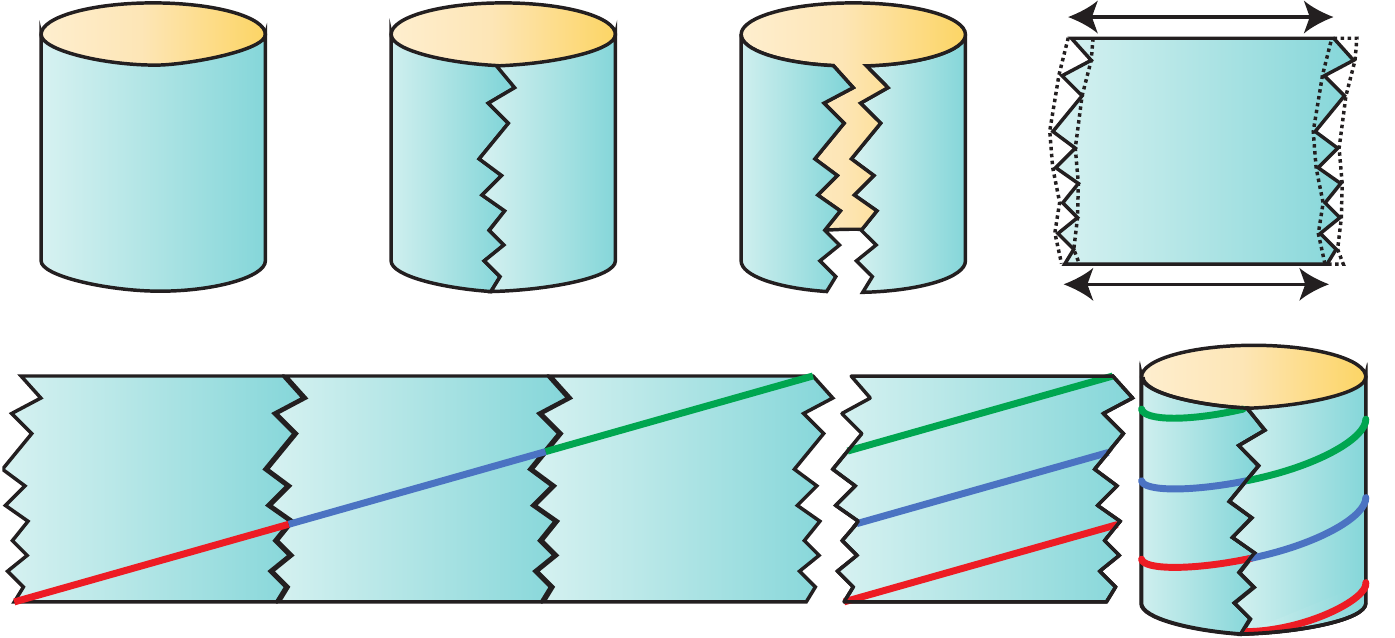}
   \caption{Drawing a spiral on a cylinder can be done by cutting the cylinder from the top boundary to the bottom one and unfolding it to the plane. We then place copies of the flattened cylinder and draw a straight line that passes from the bottom leftmost corner to the top rightmost one. Overlaying the copies and folding back to a cylinder creates a spiral, where the number of windings is equal to the number of flattened copies. }
   \label{fig:CylinderSpiral}
\end{figure}

\begin{figure}[t]
        \centering
        \includegraphics[width=1.0\columnwidth]{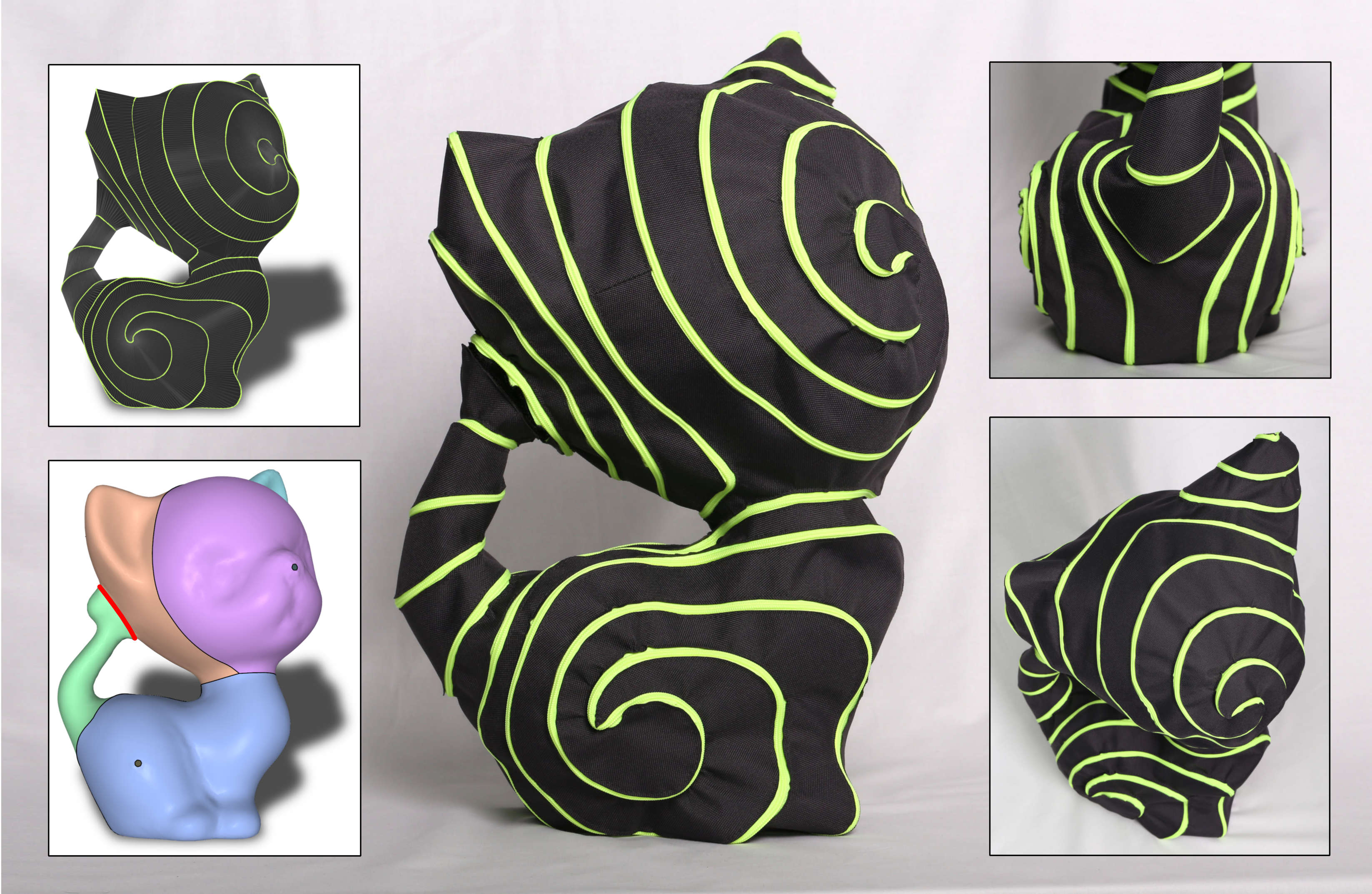}
        \caption{A zippable shape of a kitten. Since it is topologically equivalent to a torus, an additional cut is needed (bottom left inset: marked in red where the tail touches the head). Another zipper could be used to close up this cut, but we opted for using Velcro instead.}
        \label{fig:Cat}
\end{figure}

Given a mesh representing the 3D object, our goal is to generate a single, flat, possibly branching shape, which from now on we refer to simply as \emph{ribbon}, that approximates the object when ``zipped-up''. We can rigorously define zipping-up as an isometric deformation of the flat shape into a  3D shape such that the boundary exactly overlaps with itself. However, we assume that our intention is clear and avoid mathematical rigor at this point. In addition to being ``zippable'', we wish to enable some creative control by allowing the user to define where the zipper should pass and how it should be oriented or aligned.

Our method is based on the observation that it is trivial to generate a perfect spiral on a cylinder. Assume the cylinder is a topological annulus (i.e., has no caps). We can cut it from its top boundary loop to the bottom one and unfold it to a rectangular shape, where the two boundary curves become the top and bottom edges, and the cut becomes the two side edges. We then place ``copies'' of the unfolded cylinder side-by-side, and draw a straight line from the bottom corner of the leftmost edge to the top corner of the rightmost edge. Overlaying the copies on top of each other creates several disconnected, parallel line segments on the parameterization of the cylinder, and by folding it back to a cylinder, these segments transform into a perfect connected spiral. Its number of turns depends on the number of copies we made. See \figref{fig:CylinderSpiral} for an illustration.

The same approach, termed  \emph{cylindrical parameterization}, can be applied to general  cylinder-like shapes, which we colloquially continue calling ``cylinders''. We start in the same manner by cutting the shape from one boundary loop to the other. The cut shape is then mapped to the plane by a distortion minimizing parameterization with \emph{seam constraints}, which force the two sides of the cut to match like puzzle pieces. Minimizing distortion is necessary for the straight line in 2D to be mapped to a smooth and uniform spiral on the surface in 3D. The case of a more complex shape is slightly more involved: we decompose the shape into cylindrical parts and use a global parameterization scheme to smoothly map all the parts to cylinders. We discuss this in more detail in \secref{sec:Seamless}.

Once the mapping is found, we turn to designing spirals on the cylinders. The main challenge is to synchronize the spirals, such that one spiral ends where another begins, resulting in a single, long, spiraling curve on the surface. The final task is to cut the surface along the curve and remesh it such that the result is developable. It is then trivially flattened to generate the final 2D shape of the zippable ribbon.

To summarize, the design phase of our method consists of four stages:
\begin{tight_enumerate}
\item Decomposition into cylindrical parts.
\item Seamless, global parameterization of the cylinders.
\item Spiral generation.
\item Cutting along the spiral, remeshing and flattening.
\end{tight_enumerate}
See \figref{fig:teaser} for a graphical overview.

\subsection{Decomposition into cylindrical parts}\label{sec:segmentation}
\begin{figure}[b]
\centering
   \includegraphics[width=\linewidth]{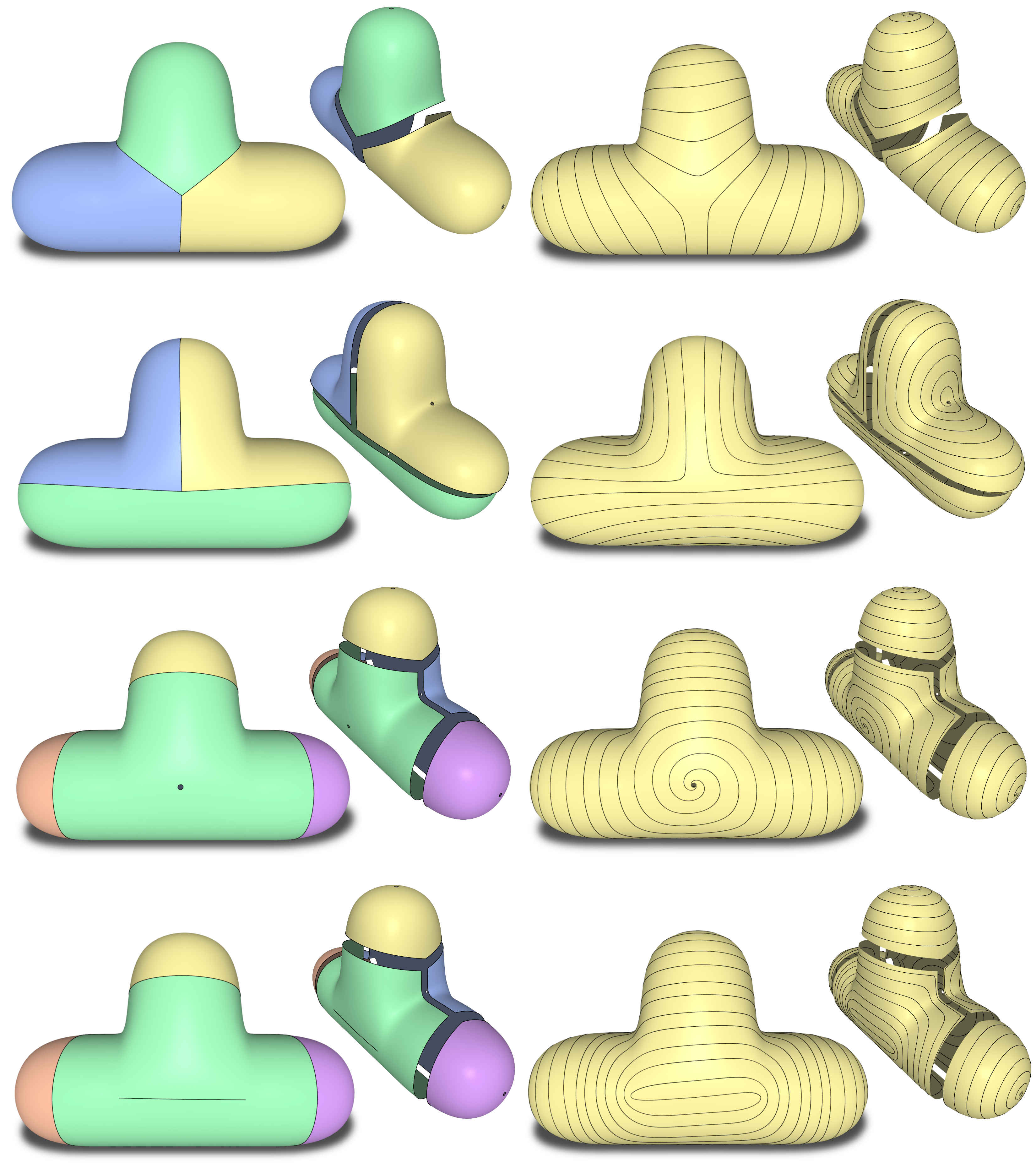}
   \caption{Different cylindrical segmentations of a T shape. Each cylinder has one transition boundary and one open boundary. Note the small holes in the middle of the colored parts. These are the open boundaries for the corresponding cylinders. In the last row we show an example of a straight curve cut, serving as the open boundary of the green cylinder. Compare the resulting spiral to the segmentation in the third row to see the effect of the curve cut.}
   \label{fig:Decomposition}
\end{figure}

We decompose the input shape $\S$ into topological cylinders $\Ci$, i.e., 2-manifolds with two boundary loops, more commonly termed \emph{annuli}. The decomposition plays a substantial role in the final appearance of the spiral, since the resulting curve is aligned to the boundaries of the cylinders. It enables a flexible interface for artistic exploration and is achieved by interactively tracing the boundaries of the segmentation using our software. Alternatively, more automatic ways of cylindrical decomposition such as \cite{zhou:15,Livesu:2017} could be applied, but we have not explored this option. Since surfaces without boundaries cannot be decomposed into topological cylinders, we also allow the user to cut the shape open and place new boundaries, e.g., small circular holes or curves on the shape that act as a single cylinder boundary (see \figref{fig:Decomposition}). We distinguish between \emph{transition} boundaries separating two adjacent cylinders and \emph{open} boundaries, which are actual boundaries of the shape. In the spiral design stage, the curve will pass through transition boundaries, \emph{spiraling in} toward an open boundary and then spiraling out toward another open boundary. This behavior is reminiscent of the Fermat spiral, which we discuss in \secref{sec:spiralDesign}.

\subsection{Seamless parameterization}\label{sec:Seamless}
\begin{figure}[h!]
   \includegraphics[width=\linewidth]{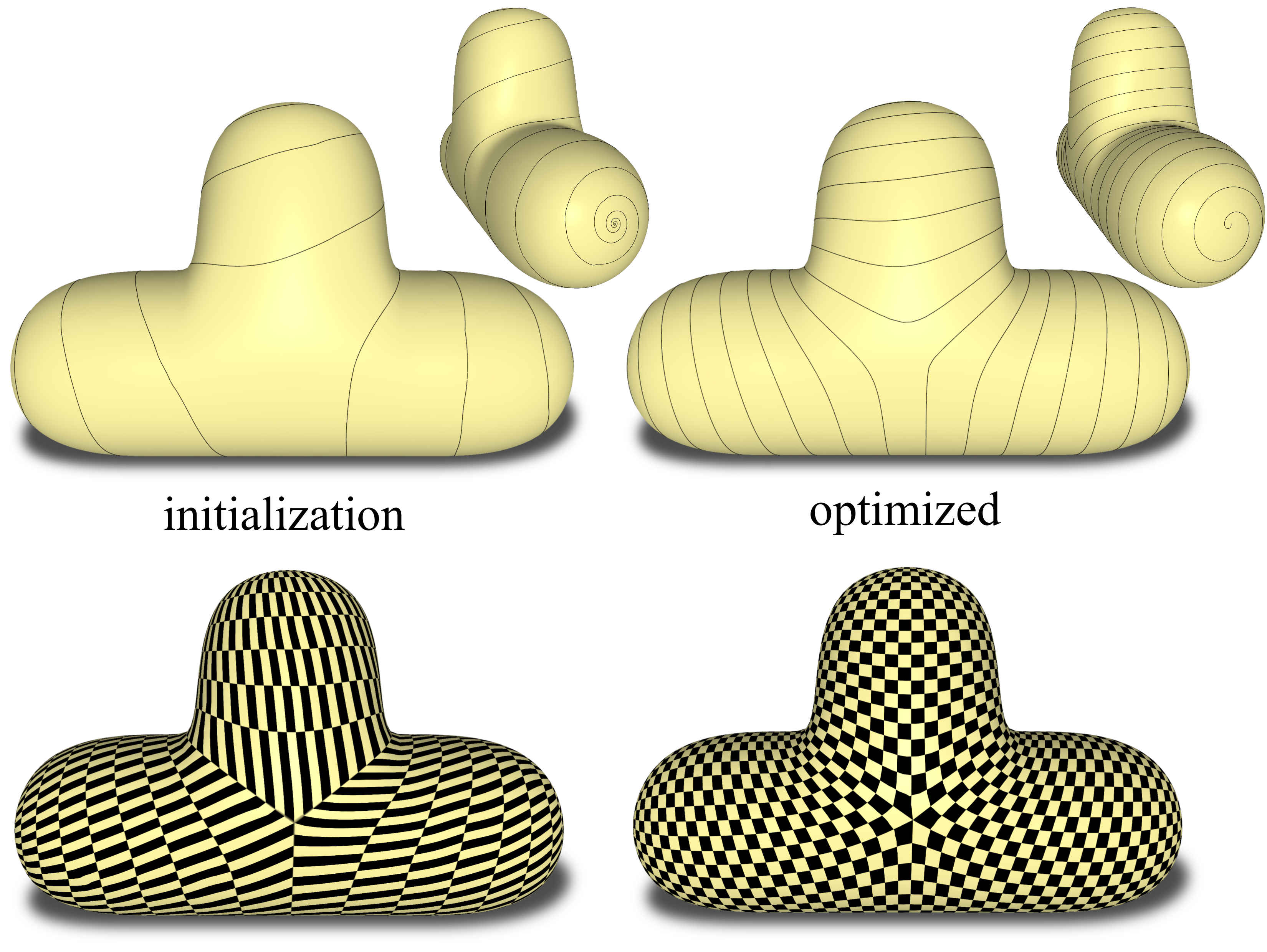}
   \caption{Comparison between the curve obtained before and after minimizing isometric distortion of the parameterization. Note that the non-optimized spirals have a much greater variation in the spacing between the windings. We show a uniform grid texture to illustrate the difference in distortion.}
   \label{fig:MinimizingDistortion}
\end{figure}

Once the cylindrical decomposition of $\S$ into parts $\Ci$ is available, we proceed to compute the parameterization. This step determines how the equally spaced, straight lines in the 2D parameter domain transform into spirals on $\S$. To obtain a spiral that maintains the even spacing on the 3D shape, the parameterization must minimize isometric distortion. Additionally, we require the parameterization to be bijective in order for the mapping from the 2D lines to the 3D curve to be well defined. \figref{fig:MinimizingDistortion} compares curves generated with the initial (i.e. suboptimal) and optimized parameterization.

We assume that $\S$ has been cut along the boundaries of the $\Ci$'s, and each cylinder is cut from one boundary to the other, analogous to the example of one cylinder (see \figref{fig:CylinderSpiral} top row). The edges and vertices along the cuts are duplicated, and we keep correspondences between the copies. We generate a seamless bijective parameterization of each $\Ci$, with seamless transitions between adjacent $\Ci$'s. We first explain the case where there is only one cylinder, and the general case of multiple cylinders  immediately follows.

\paragraph{Minimizing isometric distortion.} An isometric distortion measure quantifies the difference between a given flattening of a shape and a perfect isometry; most formulations define it as a sum of the distortions of individual triangles. In this paper we use the recently proposed \emph{symmetric Dirichlet} distortion measure; see \cite{Smith:2015,Kovalsky:2016,Rabinovich:SLIM:2017} for details.

We denote the coordinates of a vertex in the parameter domain by  $\x=(x,y)$ and stack all the coordinates in a vector $\X$. The distortion of a triangle $t$ is a function of the positions of its vertices in the plane. We denote the symmetric Dirichlet measure of triangle $t$ by $D_t(\X)$. Then the optimization problem to solve is
\begin{equation} \label{eq:OptimizationDistortion}
\argmin_\X \sum_{\text{triangle } t} A_t\, D_t(\X),
\end{equation}
where $A_t$ is the area of $t$ in the original mesh. In our work, we use a modified Newton's method \cite{Shtengel:2017} with a feasible starting point to solve this problem. Since we add several constraints in the following, we defer a detailed discussion to the end of this section.

\paragraph{Seamless cylindrical parameterization.} To map a single topological cylinder $\Ci$ to the plane with  minimal distortion in a seamless manner, it is cut to form a disk topology and then parameterized while adhering to seam constraints. The role of the seam constraints is to ensure that the parameterization is invariant to the cut \cite{MylesZ12}. In the cylinder case, the seam constraints call for each edge on one side of the seam to be a translation of its twin edge on the other side of the seam. More precisely, assume the cut contains $n$ consecutive vertices and let $\xl_j$ and $\xr_j$, $j=1,\ldots,n,$ be the two copies of each vertex in the parameterization (superscripts $\text{L},\text{R}$ stand for Left and Right). Then the cylindrical seamlessness constraints are
\begin{equation}\label{eq:CylindricalSeamlessness}
\left[\,\Cyl{\Ci}\,\right]\quad\ \xl_{j}-\xl_{j-1} = \xr_{j}-\xr_{j-1}, \quad j=2,\dots,n.
\end{equation}
We use the differential form of the seam constraints in order to avoid introducing auxiliary variables. The equivalent positional form is
$\xl_{j} = \xr_{j}+ \mathbf{t},\ j=1,\dots,n,$
where $\mathbf{t}$ is an unknown offset (the same for all vertices). For conciseness, we refer to the set of constraints in~\eqref{eq:CylindricalSeamlessness} as $\Cyl{\Ci}$ for a given $\Ci$, or $\Cyll$ in general.

\begin{figure}[t]
   \includegraphics[width=\columnwidth]{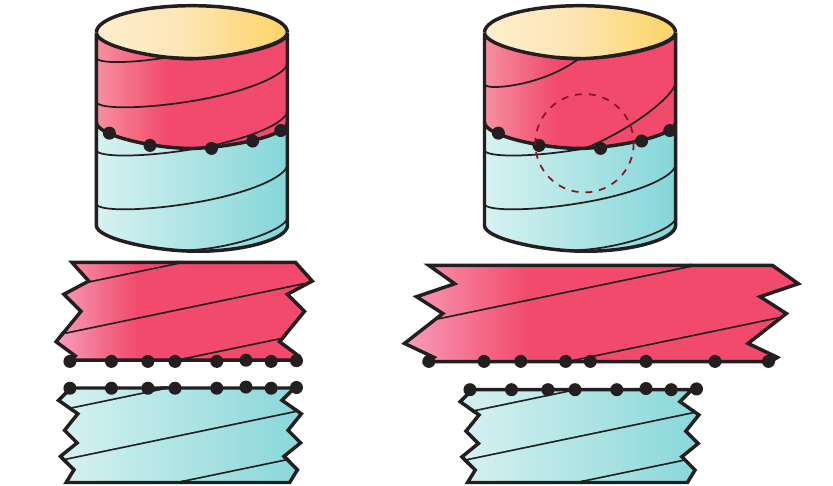}
   \caption{Inter-cylinder constraints ensure that the transitions between cylinders are smooth (left). Note the kink that appears in the curve when these constraints are missing (right). }
   \label{fig:InterCylinder}
\end{figure}

\paragraph{Cylinder boundary constraints.} In addition to the cylinder seam constraints $\Cyll$, we also require the boundary loops
\setlength{\intextsep}{7pt}%
\setlength{\columnsep}{7pt}%
\begin{wrapfigure}{r}{0.27\columnwidth}
    \centering
    \includegraphics[width=\linewidth]{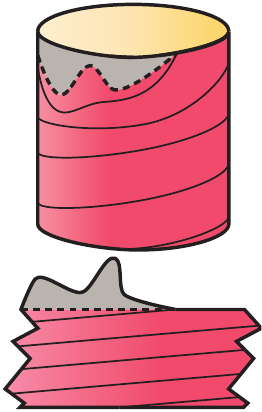}
\end{wrapfigure}
 of the cylinders to be mapped to straight lines.
This serves two purposes: First, together with the $\Cyll$, it guarantees bijectivity, and second,
it allows for a better surface coverage by  the spiral. Indeed, when the boundaries are not kept straight and allowed to ``spill out'' in the 2D domain, the spilled region is not covered by the spiral (see illustration in the inset and \figref{fig:Heart}). Due to $\Cyll$, the straight lines of the boundaries must be parallel, hence, without loss of generality, we can make them parallel to the horizontal $x$-axis. Let $\yt_{k}$, $k=1,\dots,m^\text{Top}$ and $\yb_{l}$, $l=1,\dots,m^\text{Bottom}$
be the $y$-coordinates of the vertices of the top and bottom boundaries in the parameter domain. We again use the  differential form for the straight line constraints, given by
\begin{equation}
\left[\,\Str{\Ci}\,\right] \ \  \ \
\begin{aligned}\label{eq:StraightLinesConstraints}
\yt_{k}-\yt_{k-1} &= 0, \quad k=2,\dots,m^\text{Top}\\
\yb_{l}-\yb_{l-1} &= 0, \quad l=2,\dots,m^\text{Bottom}.
\end{aligned}
\end{equation}
We denote the constraints of each $\Ci$ in \eqref{eq:StraightLinesConstraints} by $\Str{\Ci}$, and the entire set of these constraints as $\Strr$.
The constraints $\Cyll$ and $\Strr$ together already result in a nice spiral on each $\Ci$ separately. However, without dedicated treatment, there could be a visible kink in the spiral when transitioning between $\Ci$'s (see illustration in \figref{fig:InterCylinder}). Indeed, if copies of the same edge on a transition boundary are parameterized to two edges with different size and direction, then the parameterization will not appear smooth across that edge. We handle this issue in the following.

\paragraph{Inter-cylinder seamlessness.} In order for the transition between $\Ci$'s to appear smooth, we apply seam constraints on cuts between each pair of neighboring $\Ci$'s. These constraints enforce a rigid transformation between the two sides of each seam (see illustration in \figref{fig:InterCylinder}). Since we have the freedom to define the exact transformation, we choose a rotation by $\pi$. Thus, for every two neighboring $\C_P$ and $\C_Q$, we let $\x_r^P$ and $\x_r^Q$, $r=1,\dots,s,$ be the coordinates of the two copies of each vertex along the seam between $\C_P$ and $\C_Q$. Then the inter-cylinder seam constraints can be written in differential form as
\begin{equation}\label{eq:IntercylinderSeamlessness}
\left[\,\Int{\C_P}{\C_Q}\,\right]\ \ \x^P_{r}-\x^Q_{r-1} = \x^P_{r-1}-\x^Q_{r}, \quad r=2,\dots,s.
\end{equation}
We denote the constraints in \eqref{eq:IntercylinderSeamlessness} for each pair $\C_P,\C_Q$ by $\Int{\C_P}{\C_Q}$.

\paragraph{Global cylindrical parameterization.} With all the types of constraints defined, we can formulate the optimization problem for parameterizing the surface:
\begin{equation}\label{eq:UltimateOptimizationProblem}
\begin{aligned}
& \argmin_\X
& & \sum_{\text{face }t} A_t\, D_t(\X) \\
& \text{  s.t.}
& & \Cyl{\Ci}, &&\forall\, \Ci \\
&&& \Str{\Ci},  &&\forall\, \Ci \\
&&& \Int{\C_P}{\C_Q}, &&\forall\, \C_P,\C_Q\,\,\, \text{neighbors.}
\end{aligned}
\end{equation}
\paragraph{Eliminating degrees of freedom.} The constraints in \eqref{eq:UltimateOptimizationProblem} are all sparse linear homogeneous equalities, and they are overdetermined. For example, it is unnecessary to require all of the top and bottom boundaries to be on straight lines: it is sufficient to require this only for half of them, and the other half then must lie on straight lines due to the $\Intt$ constraints. Similarly, the $y$ part of \eqref{eq:UltimateOptimizationProblem} is redundant due to the $\Strr$ constraints. We remove these redundant constraints using Gaussian elimination.

\paragraph{Initialization.} Our optimization is based on Newton's method and requires a feasible starting point with no triangle flips. We use Tutte's embedding with uniform weights, which guarantees bijectivity if the boundary is convex.  We can therefore map each cylinder to a rectangle in the plane, where the top and bottom boundaries are parallel to the $x$ axis, in order to satisfy $\Strr$. We set the height of each rectangle to have the length of the cylinder boundary in order to get a more isometric initial guess. In order to satisfy $\Intt$ we require each edge of a transition boundary to have the same length in its parameterization. We can do the same for the cylinder boundary edges, which then completely determines the boundary. However, we note that we can in fact use $\Cyll$, as they are in the form of an orbifold Tutte's embedding (see \cite{Aigerman:2015}), instead of specifying vertex positions directly, which results in a slightly less distorted initial guess.

\begin{figure}[h]
   \includegraphics[width=\columnwidth]{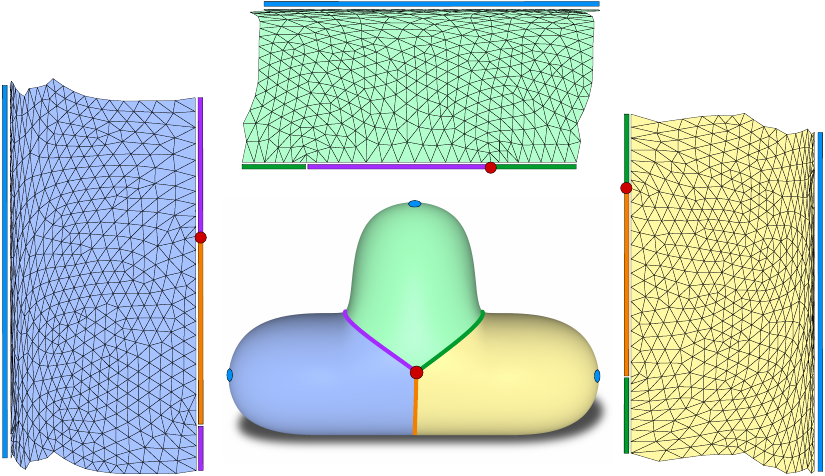}
   \caption{\label{fig:Tshapeparameterization}The parameterization of the three topological cylinders of the T shape. See Fig. \protect \ref{fig:Decomposition} for another perspective of the same segmentation. We mark copies of the same transition boundaries by matching colors, and the red dot represents one of the two points where all cylinders meet. We remark that this point has no particular significance and is only there as a visual guide. The blue sides of each flattening and the corresponding ellipses represent the open boundaries, while the unmarked sides represent the constrained cylinder seams.}
\end{figure}

\paragraph{Optimization.}
We use a modified Newton's \cite{Shtengel:2017} method with linear constraints to solve \eqref{eq:UltimateOptimizationProblem}. We use the line search method suggested in \cite{Smith:2015}, which guarantees that no triangle flips are introduced during optimization. Once the optimization converges, we continue to the next step, which is the generation of the spirals. We show an example of the parameterization of the T shape in \figref{fig:Tshapeparameterization}.

\begin{figure}[b]
        \centering
        \includegraphics[width=0.8\linewidth]{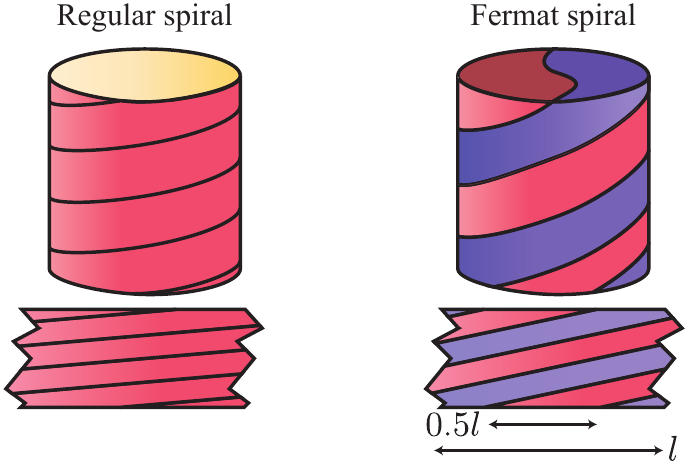}
        \caption{Illustration of a Fermat spiral on a cylinder. The cap of the cylinder on the right represents the open boundary where the center of the Fermat spiral appears.}
        \label{fig:Fermat3D}
\end{figure}

\subsection{Spiraling curve design}\label{sec:spiralDesign}

\begin{figure}[t]
        \centering
        \includegraphics[width=\columnwidth]{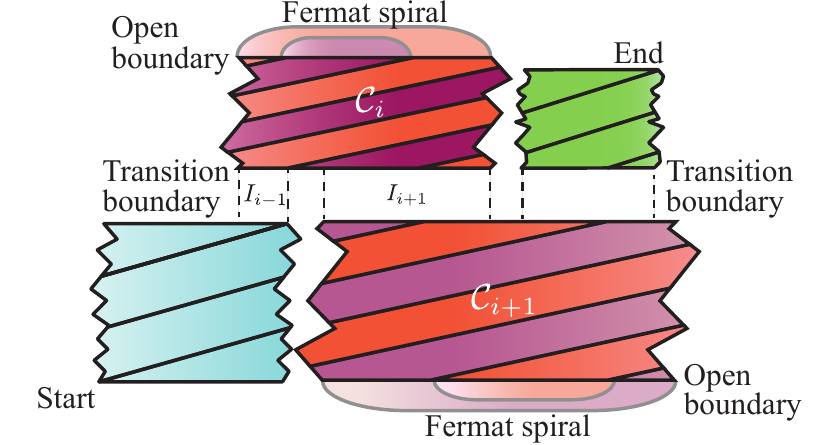}
        \caption{Illustration of a spiraling curve traversing several cylinders. We mark the interfaces of $\Ci$ by $I_{i-1}$ and $I_{i+1}$ (see \secref{sec:spiralDesign}).
        }
        \label{fig:SeveralSpirals}
\end{figure}

With the global seamless parameterization of the cylindrical decomposition available, the next stage in our algorithm is to generate a spiraling curve on the shape by drawing straight lines on the flattening and lifting them back into 3D by the inverse mapping. A spiral on a single $\Ci$ can be created as discussed in the beginning of \secref{sec:Method}, by drawing a straight line on the cylinder's parameterization. For the case of multiple $\Ci$'s, in order to obtain a single continuous curve, we must make sure that the spirals of the individual $\Ci$'s connect. We make the following simplifying assumptions:
\begin{tight_enumerate}
\item The curve starts and ends at an open boundary, passes through each transition boundary and traverses all other open boundaries via Fermat spirals.
\item Each $\Ci$ has one transition boundary and one open boundary (see~\secref{sec:segmentation}).
\item The curve enters and exits each $\Ci$ exactly once.
\end{tight_enumerate}
Note that a Fermat spiral requires drawing \emph{two} lines on a cylinder (see \figref{fig:Fermat3D}). The assumptions above mean w.l.o.g.\ that a curve must start at an open boundary in $\C_1$, then cross a transition boundary and travel to the adjacent  $\C_2$. Then touch its open boundary and leave it at another point to continue through the other transition boundary to $\C_3$. This continues until all cylinders are traversed, and the curve ends at the  open boundary of the last $\Ci$. See \figref{fig:SeveralSpirals} for an illustration.

\begin{figure}[b]
\centering
   \includegraphics[width=1.0\columnwidth]{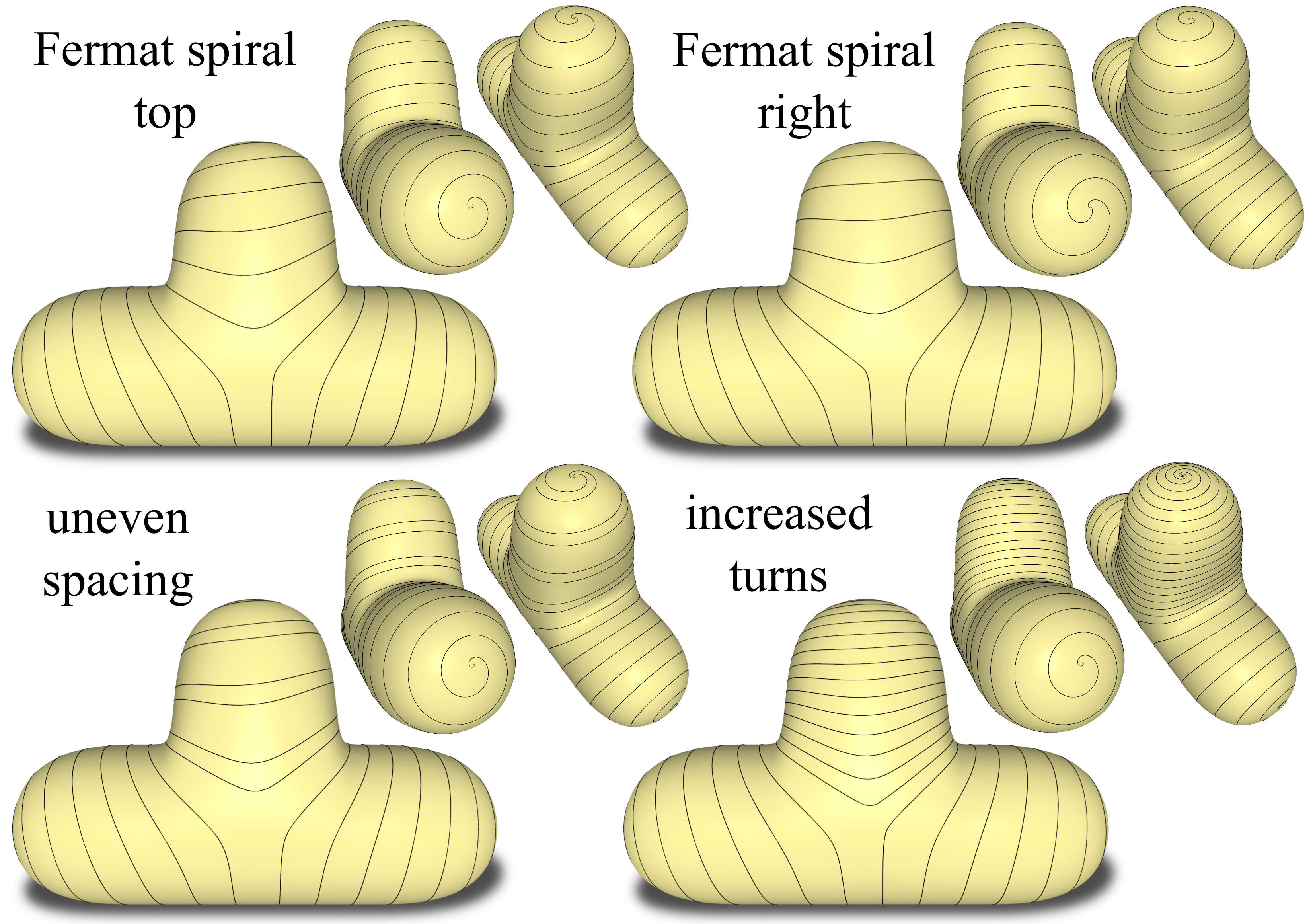}
   \caption{We show several possible spiraling curves on the T shape. The design is up to the user's artistic choices.}
   \label{fig:Options}
\end{figure}

There are several choices we let the user make. The first one is the traversal order of the $\Ci$'s. The second one is the location where the curve passes between $\Ci$ and $\C_{i+1}$. Finally, the user can prescribe the number of windings of the spiral on each $\Ci$. See \figref{fig:Options} for an example of the different choices.
These choices impact the final look of the curve and should be based mostly on artistic considerations. Having as-uniform-as-possible spacing between the windings appears to be the best for practical fabrication reasons. For a regular spiral, even spacing is ensured by the low-distortion parameterization. The Fermat spirals require a bit more thought, since two lines are needed per $\Ci$. Ideally, we would like to draw the two lines parallel, such that their copies are equally spaced. This means that the distance between their intersections with the boundary is half its width (\figref{fig:Fermat3D}). This is not always possible though, since the interfaces between $\Ci$'s might not allow it. To illustrate this,  consider a single $\Ci$, which the curve traverses after $\C_{i-1}$ and before $\C_{i+1}$ (\figref{fig:SeveralSpirals}). The $\Ci$ has an open boundary, and shares two transition boundaries with $\C_{i-1}$ and  $\C_{i+1}$ on its other boundary. The two boundaries are parameterized to two straight line segments of the same length $l$. For the open boundary, we are free to pick any two points with a distance $0.5l$ between them. Let us assume that the $x$ coordinates of the opposite boundary are the interval $I=[0,l]$. Then the interfaces with $\C_{i-1}$ and  $\C_{i+1}$ are sub-intervals $I_{i-1}, I_{i+1}\subset I$. We would like the curve to cross the boundary at two point $x_1,x_2$ such that $|x_2-x_1|=0.5l$. Clearly this is not always possible, but in practice, we did not find this to be an issue, since the user is given explicit control over these passing points. A possible future work is to solve an optimization problem that finds the best crossing points within those intervals when a problematic situation is encountered.

\subsection{Cutting, remeshing and flattening}

\begin{figure}[t]
   \centering
   \includegraphics[width=1.0\columnwidth]{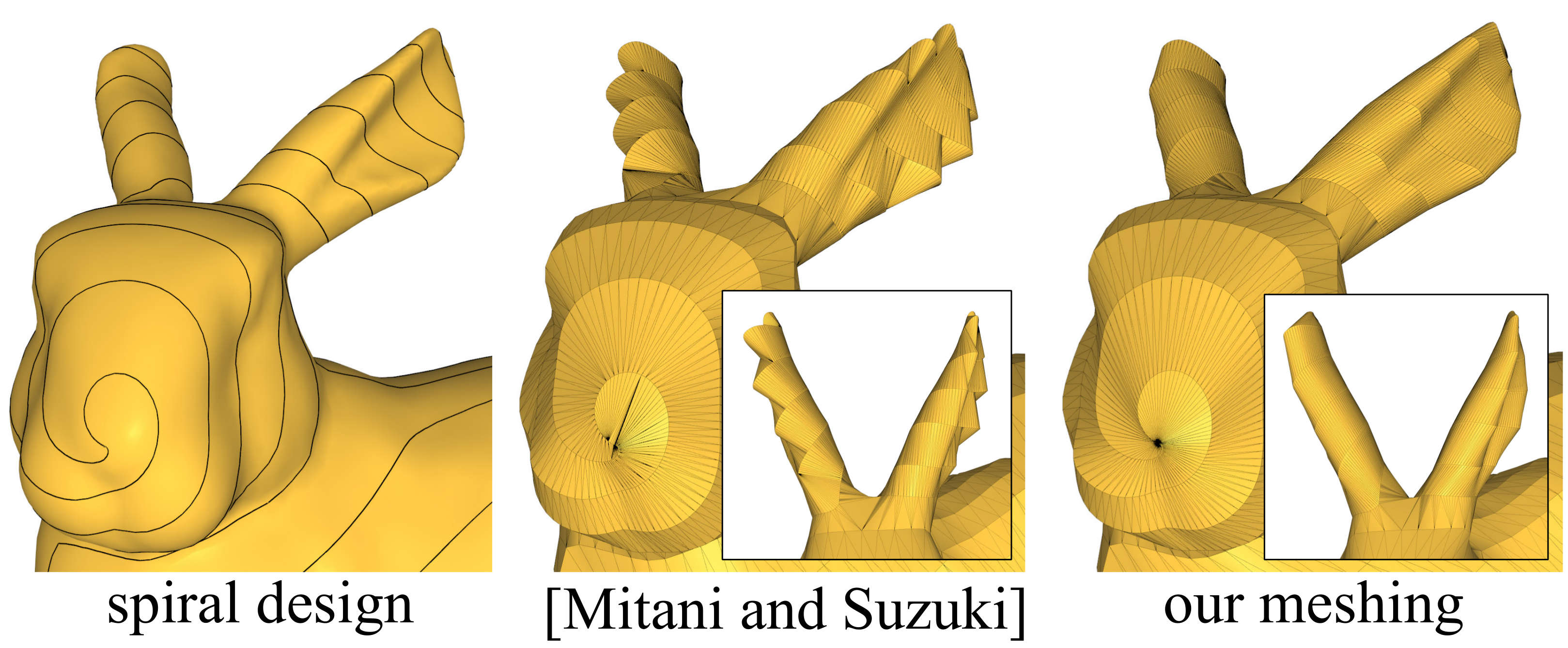}
   \caption{A comparison of our simple ribbon meshing approach to the method in {\protect\cite{MitaniS04}}: The greedy edge flipping step can generate non-optimal triangle fans (middle column), whereas ours results in a smoother and better approximation of the original surface. But it may also fail to create a good meshing (see \figref{fig:Limitation}).}
   \label{fig:remeshing2}
\end{figure}

\begin{figure}[t]
   \includegraphics[width=\columnwidth]{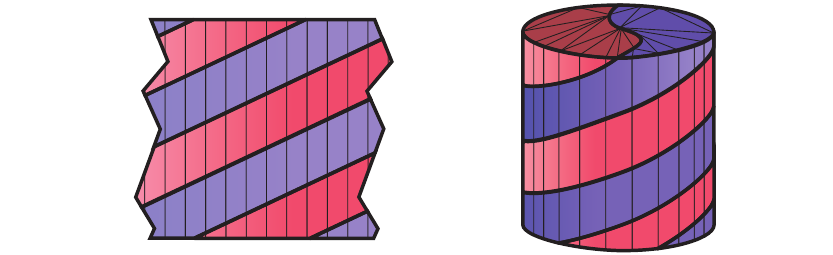}
   \caption{Illustration of a remeshing to a developable ribbon. Points on adjacent lines with the same $x$ coordinate are connected by an edge.}
   \label{fig:remeshing}
\end{figure}

\begin{figure}[h]
        \centering
        \includegraphics[width=\columnwidth]{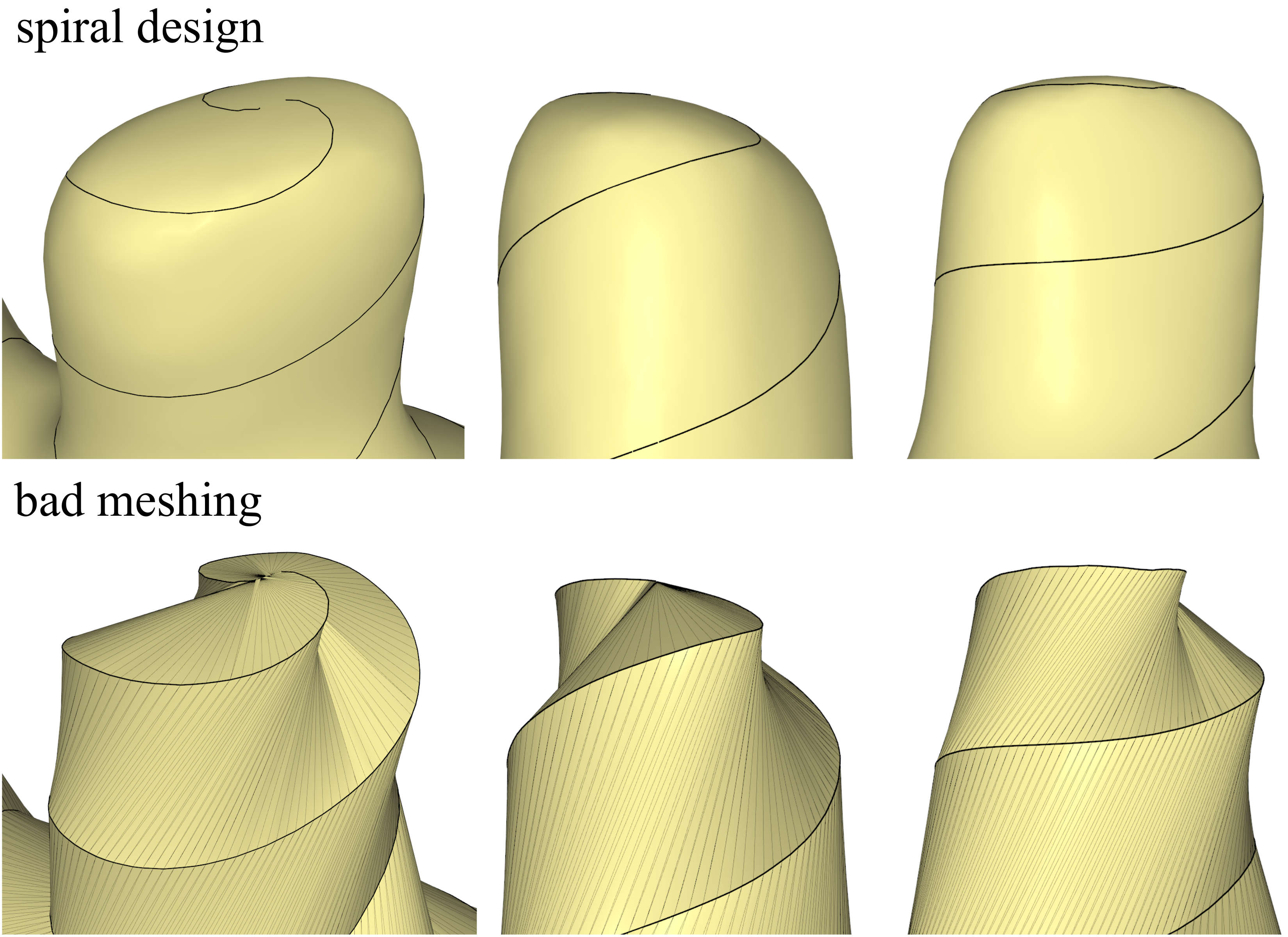}
        \caption{Our naive meshing algorithm can sometimes fail to produce a good approximation of the original surface.}
        \label{fig:Limitation}
\end{figure}

The goal of the final stage of our algorithm is to take the curve designed in the previous stage and look for a developable surface that has the curve as its boundary and approximates the shape well. This is a challenging task in general (see e.g.\  \cite{Rose:DevSurfaces:2007,Tang:2016}), but it is somewhat simpler in the discrete setting. It is well known that a triangle mesh in 3D that has no internal vertices, i.e., all its vertices are boundary vertices, is developable. Finding a developable triangulation is still a difficult problem, where the challange lies in finding a meshing that appears smooth. Mitani and Suzuki \shortcite{MitaniS04} proposed to use edge collapse and vertex removal operations until no internal vertices exist, and then to apply edge flip operations in order to improve the smoothness of the triangulation. We have implemented their method, but observed that the greedy edge flipping can introduce triangle fans near narrow curve turns (Fig. \ref{fig:remeshing2}). We instead propose a simple approach based on the parameterization we already have from previous stages. The idea is to define correspondences between points on adjacent windings of the spiral, which act as rulings of a developable surface. The simple correspondence we choose is based on the  $x$ coordinates of the lines in each $\Ci$'s parameterization (\figref{fig:remeshing}). We sample these lines uniformly and connect two samples in adjacent windings if their $x$ coordinate is the same. We use Triangle~\cite{shewchuk96b} to complete the triangulation where there is no natural correspondence, that is, around the interfaces between cylinders. Although this simple approach is not always optimal (see \figref{fig:Limitation}), we found it to be sufficient in most cases. Once the ribbon is triangulated, we can unfold it to the plane (see \figref{fig:Pipeline}). We cut the ribbon into few smaller pieces in order to utilize the cutting area better and resolve overlaps.

\section{Results and discussion}

\paragraph{Implementation details.}

We implemented all parts of our algorithm in C++, except the parameterization, which was implemented in MATLAB. The modified Newton's method converges within at most 15 iterations and and takes about 20 seconds for a mesh with 30k triangles on our Xeon CPU E5-2650 v2, 64 GB RAM machine.

\begin{figure}[h]
\centering
   \includegraphics[width=\columnwidth]{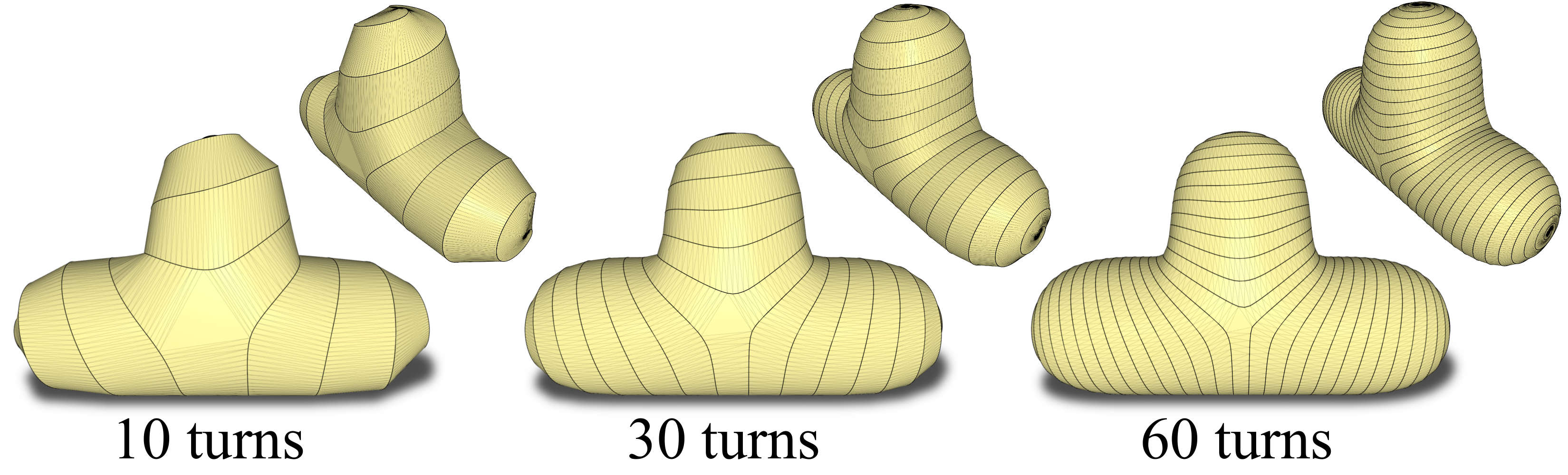}
   \caption{Running our method on the T shape with different spiral densities.}
   \label{fig:approximation}
\end{figure}

\paragraph{Approximation capabilities.} Naturally, the thinner the designed ribbon, the better its approximation power. However, this also results in a longer ribbon, prolonging the fabrication. The decision regarding the ribbon's length and width is left to the user. See \figref{fig:approximation} for different widths and approximations of the T shape.

\begin{figure}[h]
        \includegraphics[width=\columnwidth]{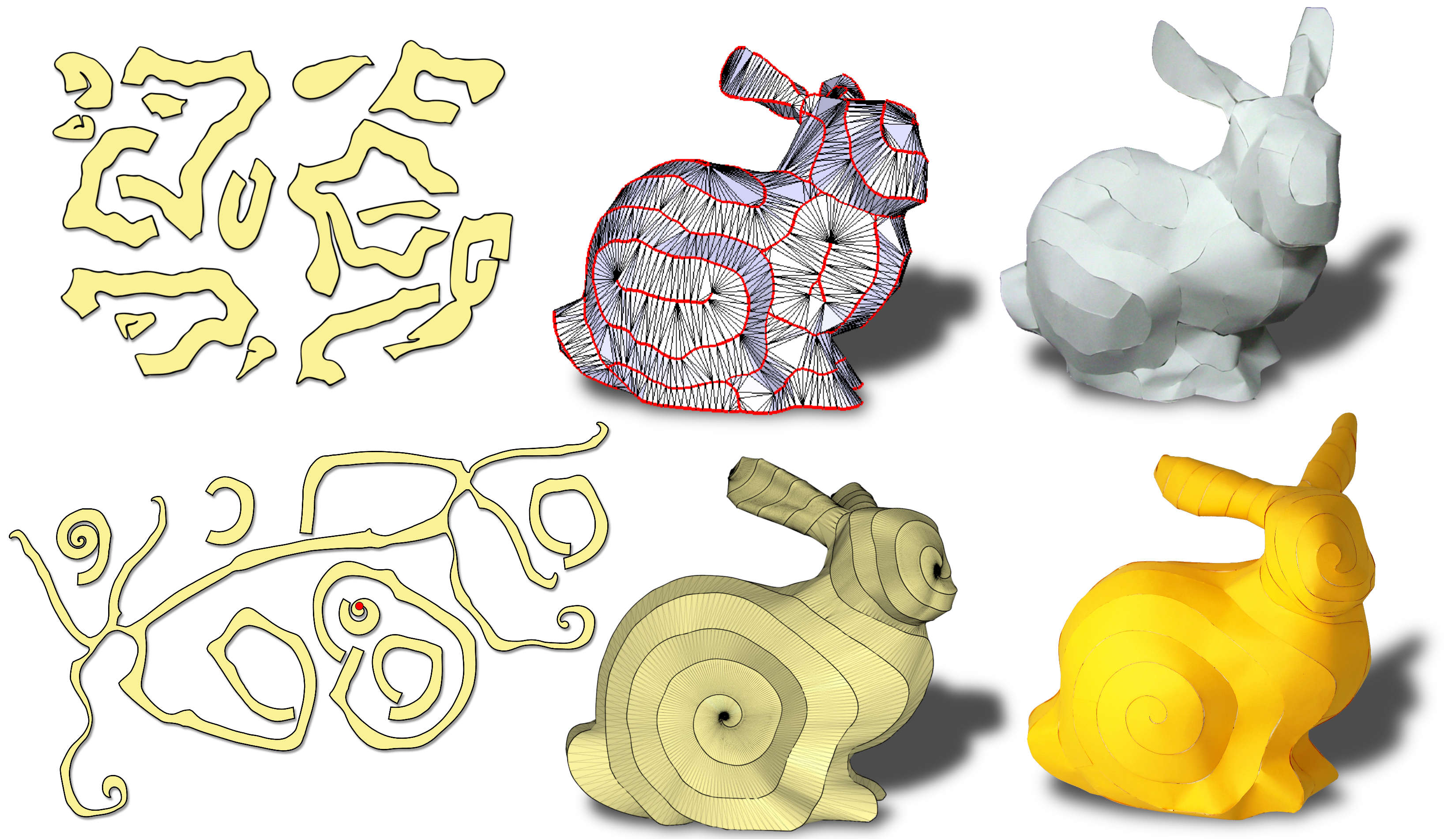}
        \caption{Our laser cut plans of the bunny with 7 pieces (bottom row) for a single developable piece that can be assembled by linearly gluing its border, starting at the designated red point, compared to {\protect\cite{MitaniS04}} (taken from their paper) with 15 pieces which require more detailed instructions to be glued together (top row).}
        \label{fig:Papercraft}
\end{figure}

\paragraph{Papercraft comparison}
Our method can also be used to create papercraft models. In general, it produces fewer initial pieces then related methods and the assembly by gluing is straight forward and does not need elaborate instructions. See \figref{fig:Papercraft} for a comparison of our bunny result fabricated from paper with the method of \cite{MitaniS04}.

\begin{figure*}[t]
        \centering
        \includegraphics[width=1\textwidth]{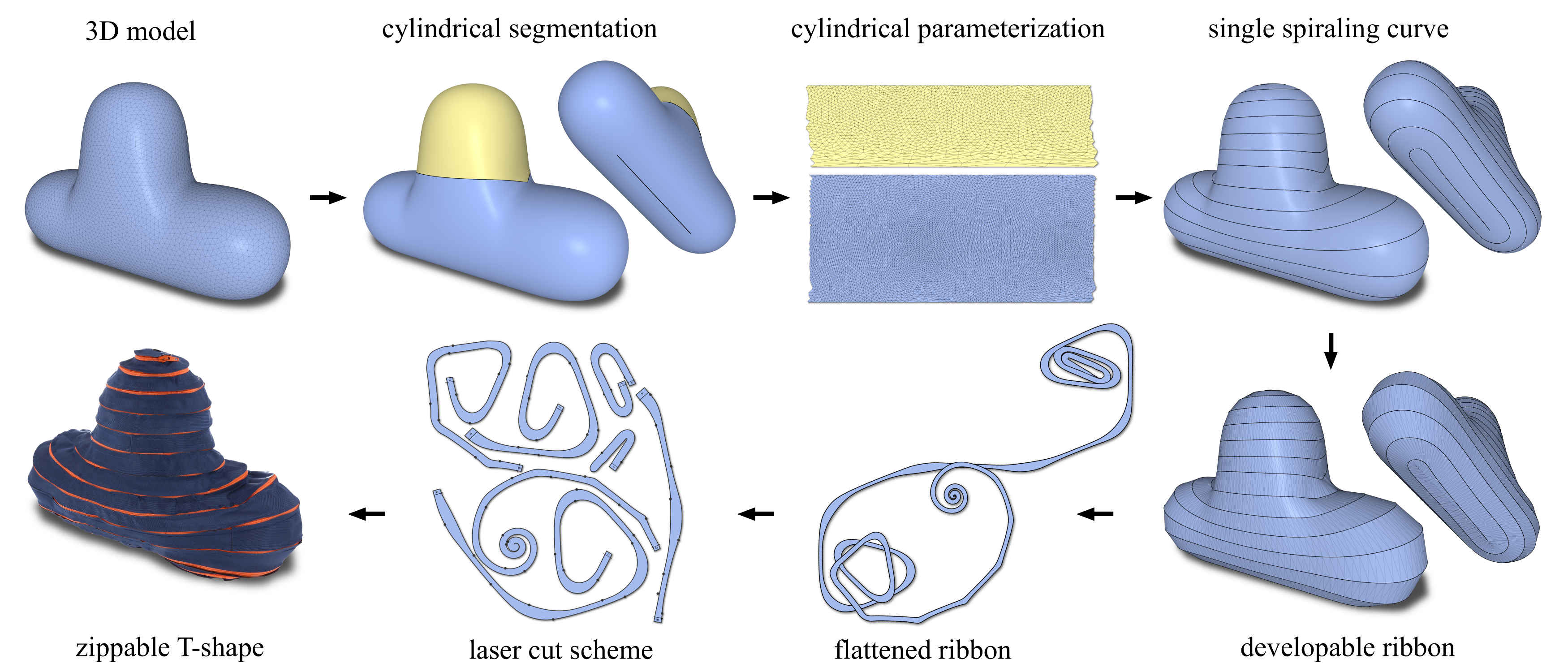}
        \caption{Overview of our pipeline. We begin by segmenting a 3D model to cylinders, followed by a global cylinder parameterization. Using the parameterization, we trace a spiraling curve on the shape. The shape is then cut along the curve and approximated by a developable ribbon. The ribbon is then unfolded to the plane. We proceed by packing the design in order to create a cutting program for a laser cutter. Finally, we cut the design from a piece of fabric and sew a zipper along its boundary. When zipped-up, the ribbon reproduces the original shape.}
        \label{fig:Pipeline}
\end{figure*}

\begin{figure}[h]
        \centering
        \includegraphics[width=1.0\columnwidth]{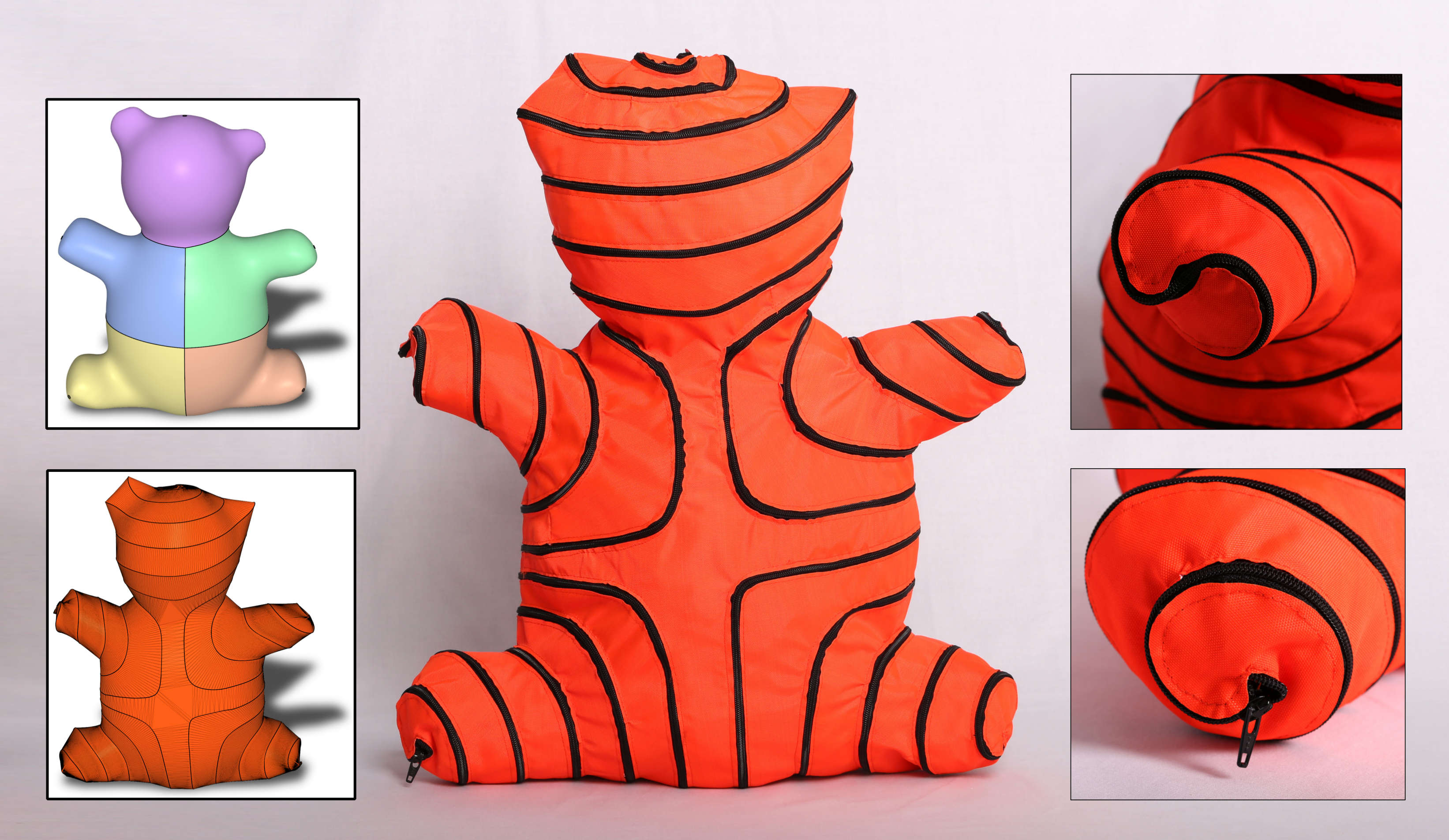}
        \caption{A zippable teddy bear with Fermat spirals for the head and the two arms.}
        \label{fig:Teddy}
        \vspace{-15pt}
\end{figure}

\begin{figure}[h]
        \centering
        \includegraphics[width=1.0\columnwidth]{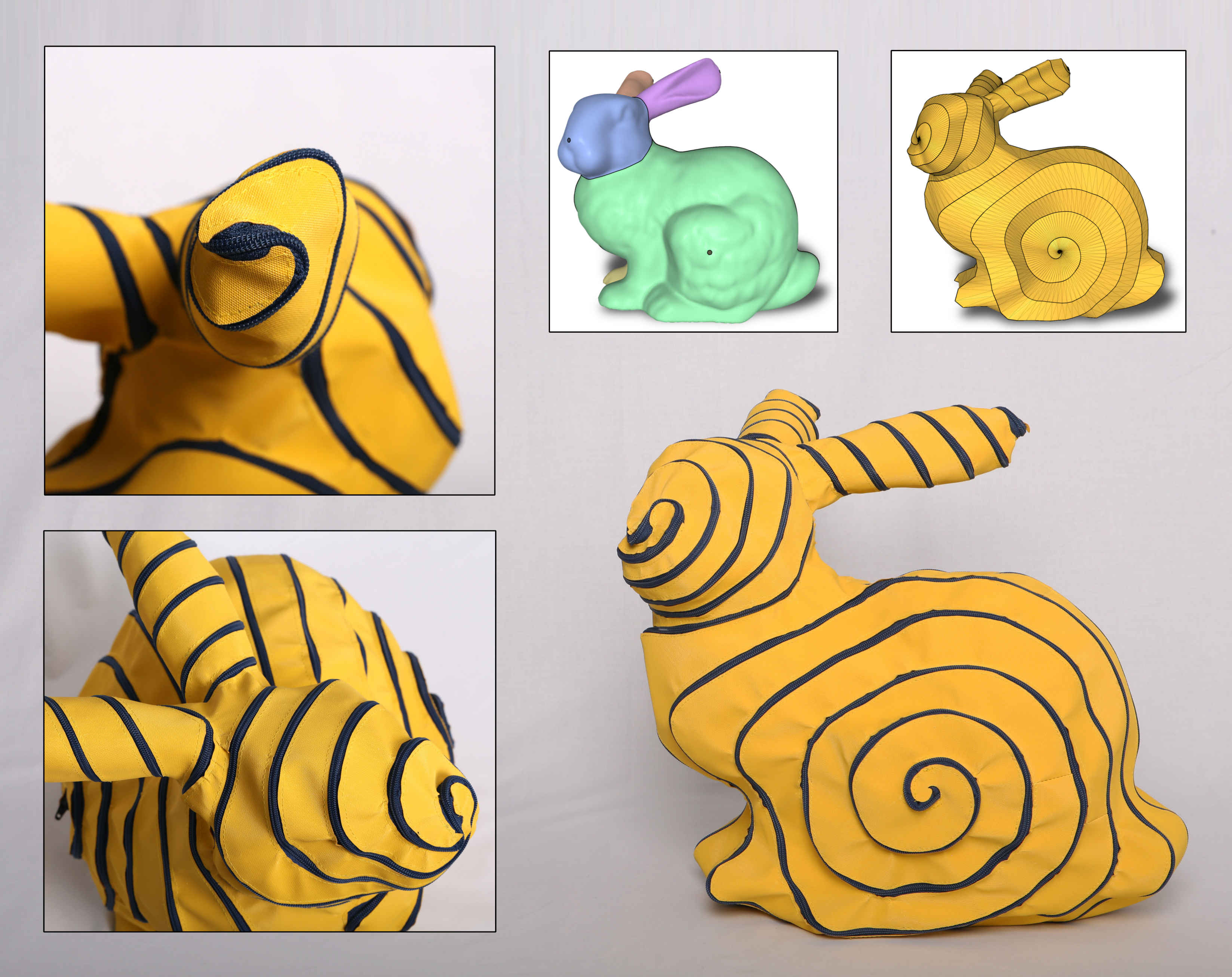}
        \caption{The Standford bunny from the teaser, zipped-up and shown from different perspectives.}
        \label{fig:Bunny}
        \vspace{-15pt}
\end{figure}

\paragraph{Fabrication process.} We fabricated six of our designs in fabric, see Figures \ref{fig:Star}, \ref{fig:Totoro}, \ref{fig:Cat}, \ref{fig:Pipeline}, \ref{fig:Teddy}, \ref{fig:Bunny}. \figref{fig:Pipeline} shows all the steps of our method with our simplest model, the T shape. All fabricated results have a zipper length of 10 meters, except the star, which has 11 meters. The models have a maximum size of around 50 cm. We asked professional tailors to do the sewing work, and it took them between 5 to 6 hours for each model. Even though they had no experience with this special kind of fabrication, there were no problems in attaching the zippers to the cut fabric pieces thanks to the linearity of the assembly method. To guarantee correct alignment of the zipper to the fabric, we place markers every 5 cm both on the zipper and along the laser cut piece. We note that care must be taken to align the teeth correctly when attaching the slider. Misalignment leads to an offset in the boundary correspondence, which results in an incorrect zip-up. However, this alignment needs to be done only once, and the markers can be taken as a guidance.

\begin{figure}[h]
        \centering
        \includegraphics[width=1.0\columnwidth]{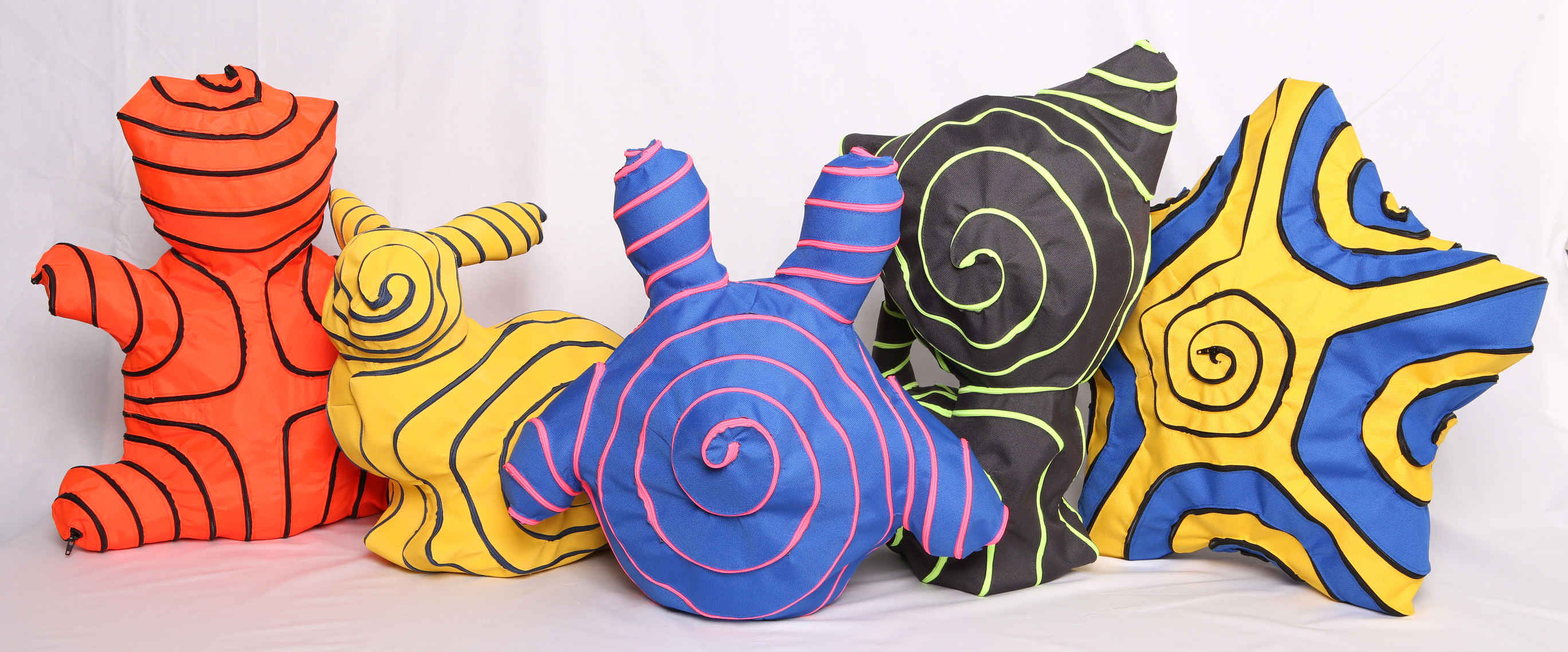}
        \caption{A display of all our fabricated results (not including the simple T shape).}
        \label{fig:OverviewResults}
\end{figure}

\begin{figure}[h]
        \centering
        \includegraphics[width=0.8\columnwidth]{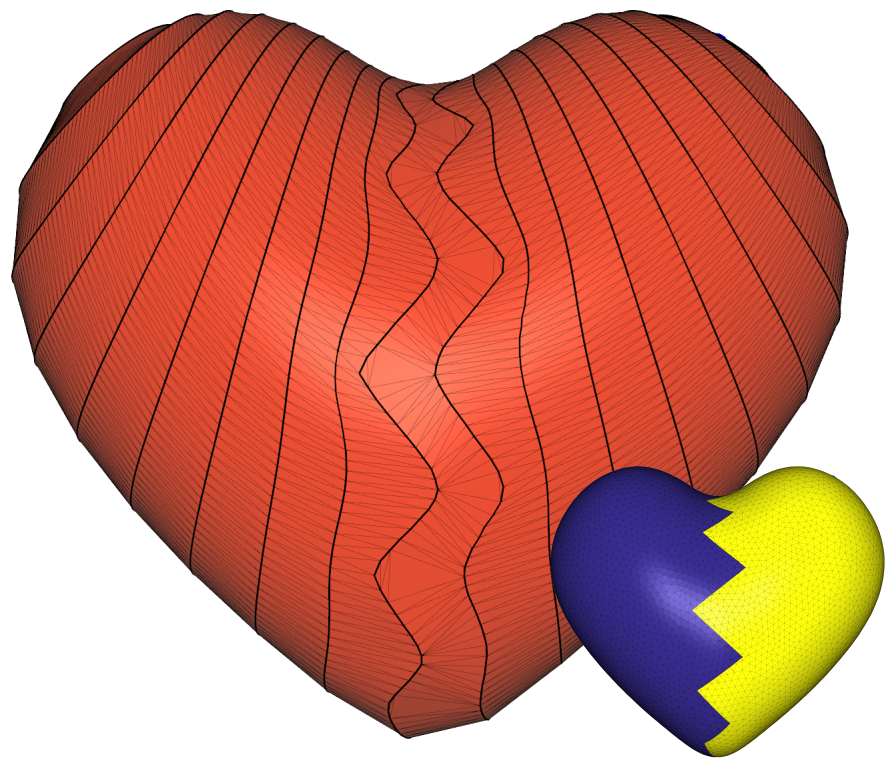}
        \caption{The heart shape is segmented in a zig-zag way to enforce a curved ribbon in the middle of it.}
        \label{fig:Heart}
\end{figure}

\begin{figure}[h]
        \centering
        \includegraphics[width=\columnwidth]{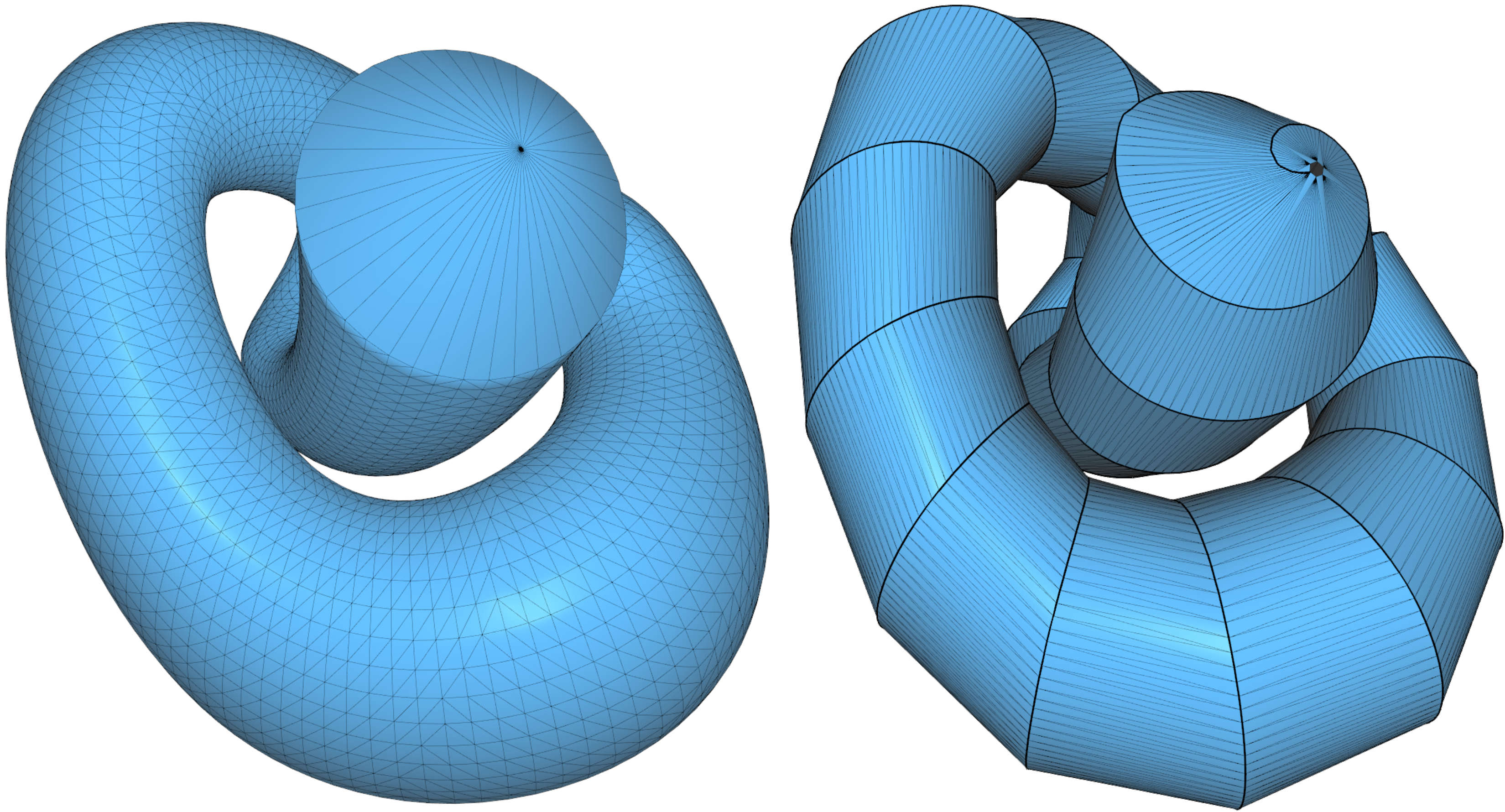}
        \caption{The developable ribbon nicely wraps around the extruded cylinder even for the long and twisty shapes. The open boundaries are at the two end points of the cylinder knot.}
        \label{fig:Knot}
        \vspace{10pt}
\end{figure}

\paragraph{Further examples.} We experimented with our system and created several additional examples. In \figref{fig:Heart} we show a heart shape with a nontrivial zig-zag segmentation. In \figref{fig:Knot} we show a result obtained on a twisting tube that changes width.

\section{Conclusion}

We presented an approach for shape representation using a single developable piece of fabric. We show several examples to demonstrate the power and generality of our approach. Currently, we do not attempt to align the zipper curve to the input shape's features. This means that regions with sharp corners may not be well represented by the assembled ribbon, unless the user manually specifies it. We plan to tackle this issue in the future by using feature detection and incorporating it into the optimization. Additionally, we are interested in targeting more global objectives, such as symmetry. Furthermore, we would like to combine the curve design with the final stage of remeshing to get a developable surface. Currently these steps are strictly decoupled, and we expect to get a better approximation by optimizing both parts simultaneously.

In this paper we mostly discussed the design part of the process, and put less emphasis on the fabrication part. In the future, we plan to explore the possibility of automating the process, extend it to other types of fabrication methods, and search for different applications. One application that we are particularly interested in is \emph{pipe cladding}, which is part of the process of insulating heated pipes with metal sheets, and is a labour intensive task. The process is similar to our zipping, but usually performed by and expert using a manual approach; our method could be potentially used to greatly simplify this task.

\section{Acknowledgements}

The authors would like to thank Isabelle von Salis for her support with the fabrication of the prototype. This work was supported in part by the ERC grant iModel (StG-2012-306877).

\bibliographystyle{acmsiggraph}
\bibliography{98-Bib}

\end{document}